\documentstyle[floats,twocolumn,aps,epsfig,prl]{revtex}

\begin{document}
\draft

\twocolumn[\hsize\textwidth\columnwidth\hsize\csname 
@twocolumnfalse\endcsname


\title{Low temperature specific heat of the heavy fermion 
superconductor PrOs$_4$Sb$_{12}$}

\author{R. Vollmer$^{(1)}$, A. Fai{\ss}t$^{(1)}$, C. 
Pfleiderer$^{(1)}$, H. v. L\"ohneysen$^{(1,2)}$, E. D. Bauer$^{(3)}$, 
P.-C. Ho$^{(3)}$ and M. B. Maple$^{(3)}$}

\address{
$^1$Physikalisches Institut, Universit\"at Karlsruhe, 
Wolfgang-Gaede-Str. 1,  D-76128 Karlsruhe, Germany\\
$^{(2)}$ Forschungszentrum Karlsruhe, Institut f\"ur 
Festk\"orperphysik, D-76021
Karlsruhe, Germany\\
$^{(3)}$ Department of Physics and Institute for Pure and Applied 
Physical Sciences,\\ 
University of California San Diego, La Jolla, CA 92093-0319, USA\\
}

\date{\today}
\maketitle

\begin{abstract}
We report the magnetic field dependence of the low temperature
specific heat of single crystals of the first Pr-based heavy fermion
superconductor PrOs$_4$Sb$_{12}$.  The low temperature specific heat
and the magnetic phase diagram inferred from specific heat,
resistivity and magnetisation provide compelling evidence of a doublet
ground state and hence superconductivity mediated by quadrupolar
fluctuations.  This establishes PrOs$_4$Sb$_{12}$ as a very strong
contender of superconductive pairing that is neither electron-phonon
nor magnetically mediated.
\end{abstract}

\pacs{PACS numbers: 74.70.Tx, 65.40.-b, 7127,+a, 75.30.Mb}
\vskip2pc]


Superconductivity mediated by a pairing potential other than a
conventional electron-phonon interaction has been the subject of a
very large number of theoretical and experimental investigations over
the decades.  In recent years, finally, intermetallic compounds have
been discovered which, together with the high-$T_{c}$ cuprates, represent
prime candidates for magnetically mediated pairing.  Surprisingly,
however, magnetically mediated pairing thus far has been considered
the only serious alternative to electron-phonon mediated pairing, 
while excitonic and polaronic mechanisms have also been proposed.

We have recently reported the discovery of the first Pr-based heavy
fermion superconductor PrOs$_{4}$Sb$_{12}$ \cite{bau02a}, for which the
nonmagnetic ground state appeared best described as a crystalline
electric field (CEF) doublet.  This in turn suggested that the heavy
electron liquid is of quadrupolar origin and that consequently quadrupolar
fluctuations mediate the superconductive pairing. 
PrOs$_{4}е$Sb$_{12}е$ hence is a candidate for being the first
material in which neither electron-phonon nor magnetic interactions
mediate the pairing.  However, a magnetic origin could not be
completely ruled out \cite{bau02a} and recent low temperature specific
heat measurements are reported to be consistent with a singlet CEF
ground state hence questioning the hypothesis of quadrupolar pairing
\cite{aok02}.

In order to settle the question of the ground state and thus the
possibility of the first example of quadrupolar mediated
superconductive pairing demands definitive establishement of (i) the
CEF level scheme, (ii) the unconventional nature of the
superconductivity, and (iii) coupling of the CEF excitations to the
conduction electrons.  Here we report specific heat
measurements of high quality single crystals of PrOs$_{4}е$Sb$_{12}е$ at
low $T$ and high magnetic field.  We show that the zero field data of
the single crystals and the magnetic phase diagram as established from
the specific heat, resistivity and magnetisation, as well as the
observation of a novel high field ground state, provide compelling
evidence of a nonmagnetic doublet CEF ground state intimately linked
to the conduction electrons.  This unambiguously establishes
quadrupolar fluctuations as the most likely pairing mechanism. 
Moreover, we observe two superconducting transitions giving evidence
of two distinct superconducting phases.

Previous experiments on pressed pellets of tiny single crystals of
PrOs$_4$Sb$_{12}$ \cite{bau02a} revealed superconductivity at
$T_{c}=1.85\rm K$ with an upper critical field $B_{c2}(T\to 0) =
2.5~\rm T$.  The superconductivity appears to involve heavy
quasiparticles with an effective mass $m^* \approx~50~m_{e}$ as
inferred from the jump in the specific heat at $T_{c}$е, the slope of
the upper critical field $B_{c2}$ near $T_{c}$, and the normal state
electronic specific heat.  

The normal state properties of PrOs$_4$Sb$_{12}$ exhibit features with
a single dominant energy scale of order several K. The low field
uniform susceptibility indicates a nonmagnetic groundstate,
characterised by a maximum at $3\rm~K$ and an enhanced zero
temperature susceptibility of 0.06 cm$^3$/mol.  Well above the maximum
a Curie--Weiss law is observed with an effective moment of $\mu_{eff} =
2.97 \mu_{B}$ and a Curie--Weiss temperature $\Theta_{\rm CW}= -16 \rm
K$.  The electrical resistivity drops monotonically from room
temperature to $T_{c}$ by nearly two orders of magnitude and displays
a shoulder at a temperature of order $7 \rm~K$, below which a very
weak $T^2$ temperature dependence is observed.  Finally, the normal
state specific heat displays a Schottky anomaly with a peak at $T^{*}
= 2.1 \rm~K$.  These data \cite{bau02a} were shown to be consistent
with a $\Gamma_{3}$ doublet ground state and $\Gamma_{5} (11 \rm K)$,
$\Gamma_{4} (130 \rm K)$ and $\Gamma_{1} (313 \rm K)$ excited states,
suggesting a quadrupolar origin of the heavy electron liquid. 
Nevertheless, a $\Gamma_{1}е$ singlet ground state and $\Gamma_{5} (6
\rm K)е$, $\Gamma_{4} (65 \rm K)$ and $\Gamma_{3} (111 \rm K)$ excited
states could not be ruled out.

Preliminary studies \cite{map02} of the low lying excitations by
inelastic neutron scattering (INS) confirm the presence of a CEF level
at 0.71\,meV (8.2\,K) and 11.5\,meV (133\,K) in excellent agreement
with previous assignments of a $\Gamma_{3}е$ ground state.  Further
evidence for such a ground state is also provided by recent non-linear
susceptibility measurements \cite{bau02c}.  Transverse-field muon spin
relaxation of the superconducting flux line lattice shows the absence
of nodal structure \cite{mac02} expected for a magnetically mediated
superconducting state, and is also consistent with a quadrupolar
origin of the superconductivity \cite{and02}.

The high quality single crystals of PrOs$_4$Sb$_{12}$ grown
from Sb flux \cite{bau01}, have very narrow superconducting
transitions in both resistivity and susceptibility and a low residual
resistivity $\rho_{0} < 5\mu \Omega \rm cm$.  The samples naturally
crystallized in a cubic shape, where the principal
crystallographic axis were confirmed to coincide with faces of the
cubes by means of Laue x-ray diffraction.  Powder x-ray diffraction
using a careful Rietveld analysis showed that the samples were single
phase with the correct Pr occupancy \cite{ho02}.  Occasional minor
impurity phases were identified to be elemental Sb and Os in
n-scattering studies \cite{map02}.  Recently, de Haas--van Alphen
oscillations were observed in single crystal samples from the same
growth, providing evidence of long charge carrier mean free paths
\cite{dHvA}.

In previous studies of pressed pellets of tiny single crystals
\cite{bau02a}, $C$ was found to be reduced by up to 50\% due to
inclusions of Sb flux.  In the present work we measured the specific
heat of an aggregate of five small single crystal pieces ($\sim 10\rm
mg$) all showing very well developed facets.  The results were
compared, where necessary, with the specific heat of a large single
crystal ($\sim 3.55\rm~mg$).  The aggregate of five single crystals displayed
the highest specific heat thus far reported, while that of the single
piece is reduced by 9\% indicating a small quantity of Sb inclusions. 
We find that $T^*$ and $T_{c}$ are sample independent.  Moreover, we
always find two distinct superconducting transitions.

\begin{figure}
\centerline{\psfig {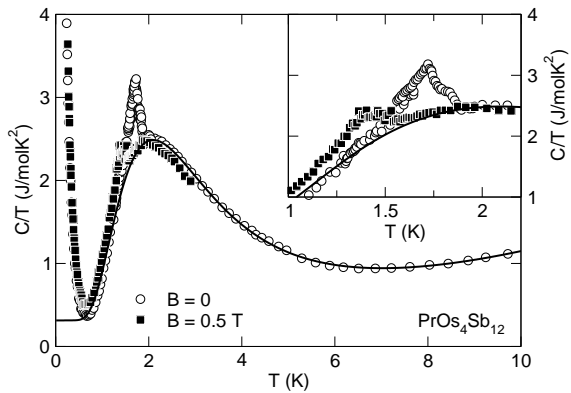}}
\caption{Specific heat $C/T$ vs.  temperature $T$ in magnetic fields
$B \parallel\,\langle 100 \rangle$ up to $0.5~\rm{T}$.  The inset displays the
specific heat near the superconducting transition, where the double 
transition can no longer be resolved at 0.5\,T.}
\label{lowfield}
\end{figure}

The specific heat was measured with a quasi-adiabatic heat pulse
technique in a $^4$He cryostat down to $1.5~\rm{K}$ in magnetic fields
up to $14~\rm T$ \cite{fai02} and in a dilution refrigerator down to
$100~\rm{mK}$ in magnetic fields up to $6.65~\rm T$ \cite{vol02}. 
Magnetic fields were applied along the principal cubic axis.  We
observe an excellent quantitative agreement to better than a few per
cent in the temperature and field range of overlap between the two
experimental set ups, which employ different thermometers and sample
holders.  We note that the specific heat of materials with large,
weakly coupled nuclear contributions as for PrOs$_{4}$Sb$_{12}$ее
display quasiadiabatic thermal relaxation times that easily exceed
several minutes.  Using a very low noise resistance bridge and
automated detection electronics we were able to keep track of these
extremely slow relaxation processes.  Our results distinctly differ
from recent work reported in reference \cite{aok02}.  However, we do
observe data consistent with those reported in \cite{aok02} when
evaluating our data using the initial $T$ versus time dependence
following a heat pulse, for which nuclear contributions are still
decoupled.

Shown in Fig.~\ref{lowfield} is the specific heat plotted as $C/T$
versus $T$ for low magnetic fields.  With decreasing temperature,
$C/T$ exhibits the well-established maximum at $T^*\approx 2.1~\rm K$,
decreases again to lower temperatures, followed by a very pronounced
upturn at the lowest $T$.  The superconducting transition at $T_c
\approx 1.85~\rm K$ is accompanied by a well developed anomaly in the
specific heat.

We focus first on the normal state specific heat.  For $B=0$ and
$T>T_c$ up to $10~\rm K$, $C/T$ may be well described as a sum of
three contributions (solid line): (i) an electronic part $C/T=\gamma
=313~\rm{mJ/molK^2}$, (ii) a phononic (cubic) part with a characteristic temperature of $\Theta = 165~\rm K$ and (iii) a Schottky anomaly due to CEF splitting of the Pr $^3$H$_4$ ground state to a $\Gamma_{3}е$ ground
state doublet and $\Gamma_{5}е$ triplet with an energy separation of
$\Delta=7.0~\rm K$.

In comparison to previous work we find a slightly reduced value of
$\gamma$.  This may be related to higher sample quality and/or the
different account of the lattice contribution, which was previously
taken as the lattice specific heat of LaOs$_4$Sb$_{12}$ ($\Theta_{D} =
304 \rm~K)$.  We did not use the lattice specific heat of
LaOs$_4$Sb$_{12}$ here, because the lanthanide contraction suggests
softer phononic rattling modes for the Pr compound than for the La
compound consistent with the value of the characteristic temperature $\Theta=165~\rm K$.

We have carefully fitted various CEF schemes for the Schottky anomaly
\cite{vol02}.  $\Delta$ can be unambiguously determined from $T^*$ of
the Schottky anomaly and is thus not a free fit parameter.  In
contrast, the absolute height of the anomaly depends sensitively on
the number of Pr ions per formula unit and the degeneracy of the
levels involved.  The data may only be accounted for by the $\Gamma_3$
doublet ground state and a $\Gamma_5$ triplet excited state, where
$\Delta = 7.0\rm~K$ corresponds very well with the INS data
\cite{map02} and the prefactor is precisely $1~\rm{Pr/f.u.}$.  For
other level splitting schemes the difference is in excess of a factor
of two.  This fully confirms the level scheme originally proposed in
\cite{bau02a}.  We also note that the electronic and lattice
contributions up to 10\,K are essentially negligible.

Evidence for unconventional superconductivity may be seen in the
specific heat anomaly (Fig.~\ref{lowfield}) in terms of two pronounced
anomalies, where the upper transition coincides with the zero
resistance and Meissner transition temperature.  The double transition
has been observed in the aggregate of five small single crystals and
also the large single piece.  It is present in all single crystalline
samples studied to date, apart from the pressed pellets \cite{bau02a},
which display a broad hump at the superconducting transition.  In all
cases for which the double transition is seen, the transition
temperatures and the size of the superconducting anomalies are in very
good {\it quantitative} agreement.  This clearly rules out sample
inhomogenities as a possible cause.  We further note that the double
transition is also seen in the thermal expansion of single crystals
\cite{oes02}.

\begin{figure}
\centerline{\psfig {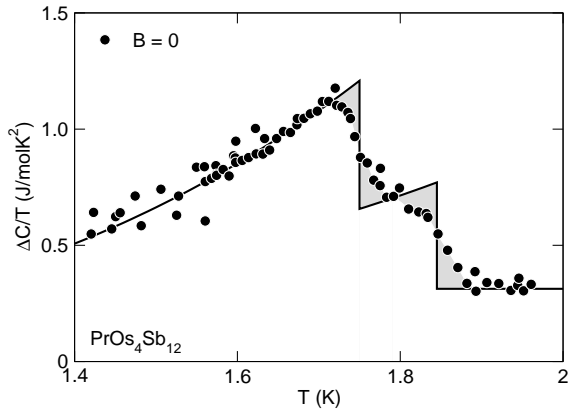}}
\caption{Electronic contribution to the specific heat $\Delta C/T$ 
vs. temperature $T$. The solid curve shows an entropy conserving 
construction as explained in the text.}
\label{Sconserve}
\end{figure}

The specific heat for $B=0$ near $T_c$ after subtraction of the phonon
and the Schottky contributions is shown as $\Delta C/T$ vs.  $T$ in
figure~\ref{Sconserve}.  The solid line represents an equal entropy
construction showing that the two transitions are of equal height. 
The two superconducting transition temperatures are estimated to
be $T_{c1}=1.75~\rm{K}$ and $T_{c2}=1.85~\rm{K}$.  The ratio $\Delta
C_{sc}/\gamma T_c\approx 3$, where $\Delta C_{sc}$ is the total height
of both superconducting jumps taken together. It exceeds the weak coupling BCS
value $\Delta C_{sc}/\gamma T_c=1.43$.  Secondly, the entropy of the total
superconducting anomaly compares with that of the normal state as
expected \cite{vol02}.

The two superconducting transitions are reminiscent of the only other
stoichiometric metal exhibiting this type of behaviour, UPt$_3$, which
displays a complex superconducting phase diagram as a function of
magnetic field \cite{has90} and pressure \cite{trap91}.  We find that
the double transition in PrOs$_4$Sb$_{12}$ may no longer be resolved
in $0.5~\rm T$, making more measurements necessary to establish the
fate of the double transition in magnetic field.

\begin{figure}
\centerline{\psfig {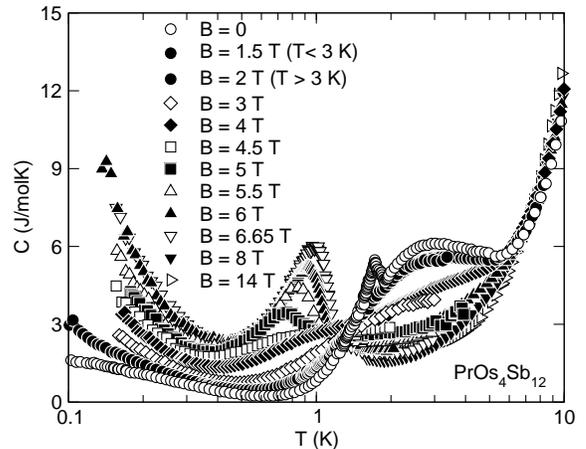}}
\caption{Specific heat $C$ vs.  temperature $T$ in magnetic fields $B
\parallel\, \langle 100 \rangle$ up to $14~\rm{T}$.  The high $T$ Schottky anomaly is
rapidly suppressed up to $14~\rm T$.  A pronounced anomaly, indicating
an ordered state stabilized in high field, is resolved at low $T$.}
\label{highfield}
\end{figure}

The specific heat $C$ as a function of $T$ in high magnetic fields up
to $14~\rm T$ is shown in Fig.~\ref{highfield}.  Data above $6.65~\rm
T$ could only be measured above $1.5~\rm K$.  The Schottky anomaly is
strongly suppressed by the magnetic field.  For $T \to 0$ a strong
increase of $C$ is observed.  Above 2\,T, data were not taken below
0.16\,K because the weakly coupled, very large nuclear contributions
could no longer be accounted for as explained above.  The low $T$
increase is consistent with the splitting of the $\Gamma_{3}е$ doublet
ground state in a magnetic field and additional contributions from
hyperfine enhanced Zeeman splitting of the Pr nuclear levels.  The
observed behaviour exceeds by a large margin that reported in
reference \cite{aok02} and clearly refutes the main argument presented
in that paper supporting a singlet ground state.

\begin{figure}
\centerline{\psfig {file=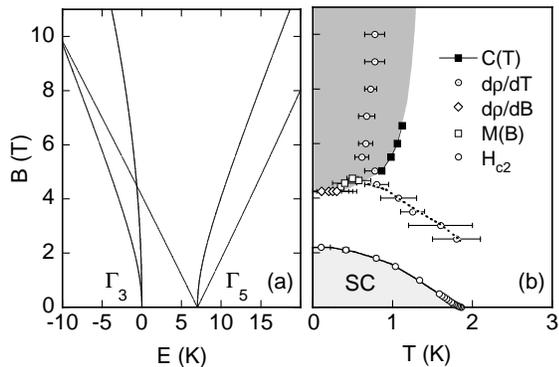,clip=,width=7.5cm}}
\caption{(a) Zeeman splitting of the $\Gamma_3$ doublet and 
$\Gamma_5$ triplet CEF levels used for fitting the zero field 
specific heat. The doublet and triplet cross at about 4.5\,T, 
suggesting a stabilisation of a different ground state at high field. 
(b) Magnetic phase diagram deduced from the specific heat $C$, 
magnetisation $M$ and derivative of the resistivity $d\rho/dT$ and $d\rho/dB$. The high field phase is clearly driven by the level crossing due to the Zeeman splitting. The zero field energy scale evident in all properties is clearly connected with the Pr$^{3+}$ $\Gamma_3$ - $\Gamma_5$ level splitting.}
\label{phase-diagram}
\end{figure}

The Zeeman splitting of the $\Gamma_{3}е$ doublet and $\Gamma_{5}е$
triplet CEF levels is shown in figure~\ref{phase-diagram}.  Here the
upper doublet and the lowest triplet level cross at $B=4.5~\rm T$
suggesting a possible change of ground state at this field.  Indeed,
the low temperature specific heat at high magnetic fields displays a
pronounced maximum for $B>4.5 ~\rm T$ that shifts from $T=0.7~\rm K$
for $B=5~\rm{T}$ to $T=1~\rm K$ for $B=6.65~\rm T$ and sharpens
considerably.  This shows the stabilisation of a new thermodynamic
ground state at high magnetic field, driven by the Zeeman split level
crossing of the $\Gamma_{3}е$ doublet and $\Gamma_{5}е$ triplet CEF. A
change of ground state is also evident from two crossing points common
to the $C(T)$ curves \cite{vol97}, one below and one above 4\,T, in
line with two fundamentally different ground states.  A quantitative
account of the high field specific heat is beyond the present work,
since the hybridisation of the f-electrons with the conduction
electrons can not be fully acocunted for.  The existence of the high
field phase nevertheless fully confirms the first excited CEF triplet
at $\Delta = 7.0 \rm~K$ and is consistent with the ground state
doublet.

The interplay of the CEF excitations with the conduction electrons is
readily evident from kinks at low fields and maxima at high fields in
the derivatives of the resistivity $d\rho/dT$ and $d\rho/dB$,
calculated from the raw data, and low $T$ magnetisation \cite{ho02}. 
When combining these features with the sharp anomaly in the specific
heat a unified phase diagram shown in figure \ref{phase-diagram} (b)
may be constructed that also agrees with features of the
magnetisation. In this phase diagram at low fields the maxima track
the calculated Zeeman splitting of the $\Gamma_3$ and the lowest
$\Gamma_{5}$ level.  This provides a natural explanation for the
energy scale of a few K at zero field seen in the resistivity,
susceptibility and specific heat.  Moreover, the very low
Kadowaki-Woods ratio $A/\gamma^2$ just above $T_c$ at zero field \cite{bau02a,map02} ($A$ is the coefficient of $T^2$ contribution to the resistivity), is boosted to a conventional value for $T \to 0$ in a magnetic field above $\sim 4\,\rm T$. Thus the CEF sensitively affects the properties of the conduction electrons, even at high magnetic fields.

The origin of the high field phase and its consistency with
the doublet ground state may be reminiscent of the properties of
PrPb$_3$ \cite{prb} and PrFe$_4$Sb$_{12}$ \cite{pfs}, both of which
stabilise antiferroquadrupolar order (AFQO) at low temperatures by
virtue of a Jahn-Teller distortion.  For PrPb$_3$, the ordering
temperature $T_{\rm AFQO}$ is shifted to higher values in magnetic
field before it collapses to zero above $\sim 7$\,T \cite{vol02}, in
qualitative agreement with the increase of the onset of the high field
phase in PrOs$_{4}$Sb$_{12}$.  The magnetic field dependence in
PrPb$_3$ is thereby driven by the Zeeman splitting of the CEF.
Although PrPb$_3$ has a $\Gamma_3$ doublet ground state and the first
excited state is a $\Gamma_4$ and not $\Gamma_5$ triplet, as for
PrOs$_4$Sb$_{12}$, AFQO at high fields is a real possibility.  A
possible suppression of the high field state and qualitative analogy
of the magnetic phase diagram with that of PrPb$_3$ may therefore
already provide compelling evidence of AFQO in magnetic fields in
PrOs$_4$Sb$_{12}$, where neutron scattering experiments at high field
are required for final proof.

In conclusion we have examined the specific heat of single crystalline
PrOs$_4$Sb$_{12}$ and found the data up to 10\,K and in high magnetic
field well explained by a $\Gamma_3$ doublet ground state and
$\Gamma_5$ excited state, consistent with the previous specific heat
in zero field on pressed pellets \cite{bau02a}.  We observe a double
superconducting transition indicating two distinct superconducting
phases, thereby highlighting the unconventional nature of the
superconductivity.  At high magnetic field, we find that a novel
groundstate is stabilised in magnetic field. The magnetic
phase diagram may tracked also by features of the electrical
resistivity.  Since the Zeeman splitting also removes fluctuations
related to the degeneracy of the doublet groundstate, these are likely
to be instrumental to the superconductive pair formation.  When taken
together, the evidence presented here establishes PrOs$_4$Sb$_{12}$ as
very strong contender for quadrupolar superconductive pairing, i.e.,
neither electron-phonon nor magnetically mediated.

Financial support by the Deutsche Forschungsgemeinschaft, U.S.
National Science Foundation, U.S. Departement of Energy and
Alexander von Humboldt Stiftung and help during the experiments by G.
Linker and F. Obermair are gratefully acknowledged.

\end{document}